\documentclass[12pt]{article}
\usepackage{epsfig}
\begin{document}
\newcommand{\beq}{\begin{equation}}
\newcommand{\eeq}{\end{equation}}
\newcommand{\sg}{\ensuremath{\sigma_p}}
\newcommand{\sm}{\ensuremath{\sigma_c}}
\newcommand{\lf}{\ensuremath{\langle L_f \rangle}}
\newcommand{\la}{\ensuremath{\langle L_a \rangle}}
\newcommand{\tf}{\ensuremath{{\rm Tr}_f}}
\newcommand{\nt}{\ensuremath{N_\tau}}
\newcommand{\ns}{\ensuremath{N_\sigma}}
\newcommand{\f}{\ensuremath{\beta_f}}
\newcommand{\ba}{\ensuremath{\beta_a}}
\newcommand{\bv}{\ensuremath{\beta_v}}
\newcommand{\bt}{\ensuremath{\beta}}
\newcommand{\bvc}{\ensuremath{\beta_{vc}}}
\newcommand{\lm}{\ensuremath{\lambda}}
\begin{titlepage}
\begin{flushright}
hep-lat/9901006 \\
TIFR/TH/98-39 \\
\end{flushright}
\vskip2cm
\begin{center}
{\Large{\bf The Deconfinement Transition in SO(3) Gauge Theory}} \\[1cm]
{\bf Saumen Datta\footnote{E-mail:saumen@theory.tifr.res.in}
 and R.V. Gavai\footnote{E-mail:gavai@theory.tifr.res.in}} \\[2mm]
{\em Theoretical Physics Group, \\
Tata Institute of Fundamental Research, \\
Homi Bhabha Road, Mumbai 400005, India} \\[1cm]
\end{center}
\begin{abstract}
The $SO(3)$ lattice gauge theory with a Villain form of action
was investigated by Monte Carlo techniques on asymmetric lattices with
\nt\ = 2 and 4, where \nt\ is the number of sites in the temporal
extent. Unlike the results for higher \nt\ , only one 
transition of second order was found for \nt\ = 2 . 
An extended action with an irrelevant term to suppress $Z_2$ monopoles
enabled us to get a better view of the deconfinement transition as the
effects of bulk transition could be suppressed as well.
Although the action has no global $Z_2$-symmetry for the $SO(3)$ theory, 
unlike the $SU(2)$ theory at finite temperature, our study revealed a
second order deconfinement transition, with properties similar to the
deconfinement transition of $SU(2)$.
\end{abstract}
\end{titlepage}

\section{INTRODUCTION}
\label{intro}

The most convenient way of regularizing gauge theories for nonperturbative 
studies is by discretizing the underlying space-time on a lattice. The
gauge variables can then be chosen as suitable representations of the 
gauge group residing on the links of the lattice.

A natural choice for the lattice action is the Wilson
action \cite{wil}, in which the gauge variables are in the fundamental
representation of the gauge group. For the SU(2) gauge theory, the action
is simply
\beq
S_W = {\f \over 2} ~\sum_p ~\tf (U_p)
\label{eq.wlac}
\eeq
where $U_p$ is the product of link variables around a plaquette, and
the subscript $f$ denotes the trace is to be taken in the fundamental
representation of the gauge group. The fundamental representation is
particularly convenient when one includes fermions, since the fermions
are normally taken in this representation. The Wilson action has been
widely studied in the literature, both for zero-temperature and finite
temperature studies. Studies of gauge theories at finite temperature 
can be performed by taking a lattice with a finite temporal extent 
equal to the inverse temperature, and imposing periodic boundary 
conditions on the gauge variables in the temporal direction.
One interesting prediction that has emerged from
such studies is that, at a sufficiently high temperature, the theory
undergoes a deconfinement transition. The finite temperature gauge theory
has a global symmetry corresponding to the center of the gauge group.
The deconfinement transition is generally described by the
breaking of this center symmetry. The Polyakov loop operator,
\beq
L_f (\vec{r}) = \tf ~\prod_{i=1}^{\nt} U_{\tau} (\vec{r},i),
\label{eq.polf}
\eeq
transforms non-trivially under this symmetry and therefore acts as an 
order parameter for the deconfinement transition. Here \nt\ is the
number of lattice points in the temporal direction.

The Wilson action (\ref{eq.wlac}) is by no means the only
choice for the discretized action. Since the continuum theory is at
a critical point of the lattice regularized theory, one expects a
large class of actions, differing only by irrelevant terms, to
correspond to the same continuum physics. In particular, one can take
the trace in (\ref{eq.wlac}) in any representation of the gauge
group. Since the continuum theory is defined only in terms of the
underlying algebra, the choice of the representation for the link 
variables should be completely irrelevant for it.

One particularly interesting variant of (\ref{eq.wlac}) is when the
trace is taken in the adjoint representation. The gauge variables in 
adjoint representation are blind to the center of
the group. The center symmetry is therefore absent and \lf\ is
identically zero. A study of the deconfinement transition for this form of
the action is therefore of particular interest \cite{smi}.  

Finite temperature $SO(3)$ gauge theory was studied in refs. \cite{sri,
sau1} with the Wilson form for the action, and in ref. \cite{sau2}
using a Villain form \cite{hal1}. A natural analog of \lf\ 
was taken to be \la\, where the adjoint Polyakov loop $L_a$ is defined as
\beq
L_a (\vec{r}) = {\rm Tr}_a ~\prod_{i=1}^{\nt} U_{\tau} (\vec{r},i).
\label{eq.pol}
\eeq
While \la\ is not an order parameter for the deconfinement
transition as the screening of an adjoint quark renders it to be
non-zero in the confined phase, it is known to
behave quite similarly as \lf\ across the $SU(2)$ deconfinement
transition. It was found in refs.\cite{sri,sau1,sau2} that the finite
temperature theory shows a discontinuous transition. $\la \simeq 0$ till
the transition point, where it rises sharply, indicating a 
deconfining nature of the transition. The transition point, however, was 
coincident with the known bulk transition point for $SO(3)$
\cite{bha}. Moreover, it did not shift with increasing temporal extent
of the lattice \cite{sau1,sau2}, which is both unlike the deconfinement
transition in the $SU(2)$ theory and characteristic of a bulk
transition.

These results appear to be similar to those obtained in refs. \cite{gav,
mat} in finite temperature studies with the Bhanot-Creutz action \cite{bha}
\beq
S = \sum_p \left[ \f \left(1 - {1 \over N} \tf U_p \right) + \ba \left(
1 - {1 \over 3} {\rm Tr}_a U_p \right) \right].
\label{eq.bcac}
\eeq
There too, a first order deconfinement line, as characterized by the
order parameter \lf\ , was found to be coincident with the previously
known first order bulk transition line for a range of large \ba\ .  
Moreover, the shift in the transition coupling with changing \nt\ was
found to be negligibly small although it increased progressively as 
\ba\ was decreased. 

Both for the intrinsic importance of the study of the deconfinement
transition for the $SO(3)$ gauge theory, and in order to have a better
understanding of the interplay between the bulk and deconfinement
transitions, we thought it necessary to conduct further detailed studies of
the $SO(3)$ gauge theory at finite temperature. To decouple the two
transitions, we 1) employed \nt\ = 2 lattices in our simulations
(a sizeable shift in the transition point was seen for \nt\ = 2
lattices in ref. \cite{mat} for the Bhanot-Creutz action) and 2) have 
used extended actions \cite{hal2} to suppress the bulk transition so
as to get a clearer view of the deconfinement transition.  As in ref. 
\cite{sau2}, we have used the Villain form of the $SO(3)$ action.  
The plan of our paper is as follows: in Sec. \ref{sc.nt2}, we present 
results for the phase transition for the \nt\ = 2  and 4 lattices. 
In Sec. \ref{sc.mono}, the phase transition for the extended action 
is discussed. The last section contains a summary of our results and 
their discussion.

\section{PHASE TRANSITION FOR THE \\ VILLAIN ACTION}
\label{sc.nt2}

\subsection{The Action and its Properties}
\label{ssc.nt2ac}

In the Villain form of the action for the $SO(3)$ theory from
ref. \cite{hal1}, one introduces auxiliary $Z_2$ variables \sg\ associated 
with the elementary plaquettes, besides the usual gauge variables in the 
fundamental representation of the gauge group $SU(2)$. 
The $SO(3)$ theory is defined by the partition function
\beq
Z = \sum_{\lbrace \sg=\pm 1\rbrace} \int {\mathcal{D}} U {\rm exp}
\bigl(S(U,\sigma)\bigr)
\label{eq.vlth}
\eeq
where the action $S$ is given by
\beq
S(U, \sigma) = {\bv \over 2} \sum_p {\rm Tr}_f (U_p). ~\sg .
\label{eq.vlac}
\eeq
On performing the summation over the \sg\ variables, one gets the
gauge field effective action 
\beq
S_{\rm eff} (U) = \sum_p {\rm ln} ~{\rm cosh} \left({\bv \over 2} {\rm
Tr}_f U_p \right)
\label{eq.effac}
\eeq
 which is clearly blind to the center symmetry. 
The character expansion of $e^{S_{\rm eff}}$ has contributions only 
from the integer representations of $SU(2)$ \cite{hal2}. Explicitly, 
the link variables have the local $Z_2$ symmetry 
\beq
U_l \to - U_l, ~\sg \to - \sg ~\forall p \ni \hat{\partial} l.
\label{eq.center}
\eeq
The connection to the $SU(2)$ gauge theory with Wilson action in
the weak coupling limit is clear: as $\bv \to \infty$, the
configurations that contribute are
\beq
U_l = \sigma_l e^{i g A}, ~~\sg = \prod_{l \in \partial p} \sigma_l
\label{eq.weak}
\eeq
where $\sigma_l$ are $Z_2$ variables situated on the links. Then a 
change of variable $U_l \to U_l \sigma_l$ reproduces the $SU(2)$ gauge 
theory with Wilson action.

The theory (\ref{eq.vlth}) is known to have a bulk transition at $\bv
\sim 4.45$ \cite{hal1,sau2}. This bulk transition has been explained by the
condensate of $Z_2$ monopoles \cite{hal1}. The two phases differ in
the value of the monopole density, defined by \cite{hal2}
\begin{eqnarray}
\label{eq.dnst}
M &=& 1 - \langle {1 \over N_c} \sum_c \sm \rangle, \\
\label{eq.mono}
\sm &=& \prod_{p \in \partial c} \sg
\end{eqnarray}
where $c$ denote the elementary cubes in the lattice and $N_c$ is the
number of such cubes. In the strong coupling phase, \[ M = 1 - 4
\bt_{\frac{1}{2}}^6 \] where the character expansion coefficients
$\bt_j$ are given by
\beq
\bt_j = {I_{2 j+1}(\bv) \over I_1 (\bv)},
\label{eq.strong}
\eeq 
whereas in the weak coupling region, $M = 0$ up to exponential corrections. 
At the bulk phase transition point, $M$ jumps from its strong coupling
value to the weak coupling value.

\subsection{Phase Transition at Finite Temperature}
\label{ssc.nt2rs} 

Numerical investigations of the finite temperature transition for this
theory have been carried out in ref.\cite{sau2}. By monitoring the
plaquette variable, $P= \sg \tf (U_p) $, and \la\ (Eq. (\ref{eq.pol})), 
for lattices with \nt\ ranging from 4 to 8, it was found that the 
theory has only one, strong first order, transition.  The transition point 
was found to be $\bv \sim 4.45$ for all the lattices, with no perceptible 
change in \bvc\ .  While this is typical of a bulk transition,
the behavior of \la\ was indicative of a deconfining nature of the
transition : it jumped across the transition, being $\simeq 0$
before the transition.  One possible way to explain the curious mixture
of the bulk and deconfining characteristics of the transition is to
hypothesize a shielding of the deconfinement transition
by a sufficiently strong first order bulk transition.  However,
since the deconfinement transition point is
expected to shift with a change in \nt\ , such a shielding will 
only be possible over a finite range of \nt\ ; outside that range the two
transitions have to separate out. Such a separation was not seen in
ref. \cite{sau2} for \nt\ = 4 to 8. Guided by the results of
ref. \cite{mat}, here we study \nt\ = 2 to explore the lower end of the
range, as the shift in the deconfinement transition point is expected to be 
larger in going from \nt\ = 4 to 2.

The transition point was indeed found to shift to $\bv \sim 4.16$ for
\nt\ = 2. The nature of the transition, on the other hand, was found 
to be different. While a two-peak structure was found at the critical 
point, it did not sharpen with an increase in the spatial size of the 
lattice. We carried numerical simulations for lattices with spatial
size \ns\ = 6, 8, 10, 12, 16 and 24 (details of the simulation are
given in Appendix \ref{app.sim}). Figure \ref{fg.hist} shows the 
distribution of the plaquette and $L_a$ for these lattices at \bvc\ . 
A two-peak structure is visible in all of them. However, the peaks move 
closer to each other as \ns\ increases and the valley in between them 
becomes shallower.  This is, of course, opposite to what one expects 
for a first order transition, and suggests either a second order transition 
or a crossover in the thermodynamic limit. From Fig. \ref{fg.hist} one can 
calculate the free energy difference between the maxima and the minima
for a given spatial volume, by using the spectral density method \cite{fer1} 
to extrapolate the histograms to a $\beta$-value such that the 
two peaks are of equal height. For a first order transition, this difference 
$\triangle F $ is expected to increase with lattice size, while a plateau
should be reached for a second order transition \cite{lee}.  The estimates 
of $\triangle F$, shown in table \ref{tb.lk}, clearly rule out a first order 
transition.

\begin{table}
\caption{Free energy difference for Lee-Kosterlitz analysis.}
\begin{center}
\begin{tabular}{|c||c|c|c|c|c|c|}
\hline
$\ns$ &6 &8 &10 &12 &16 &24 \\
\hline
$\triangle F$ &.38(2) &.52(2) &.55(2) &.48(3) &.58(5) &.57(10) \\
\hline
\end{tabular}
\end{center}
\label{tb.lk}\end{table} 

\begin{figure}[htbp]
\begin{center}
\epsfig{figure=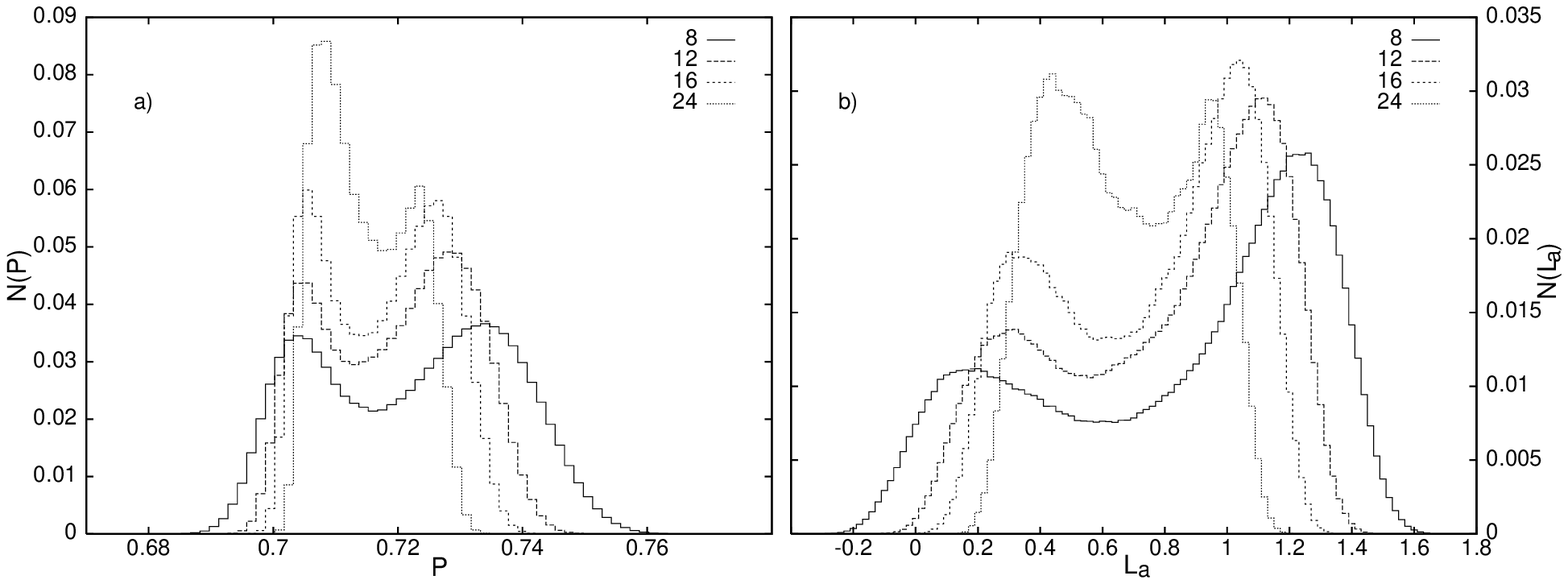,height=6cm,width=12cm}
\caption{Distribution of a) plaquette and b) $L_a$ for 
$\ns^3 \times 2$ lattices with \ns\ = 8, 12, 16, 24.} 
\label{fg.hist}
\end{center}\end{figure}

\begin{figure}[htbp]
\begin{center}
\epsfig{height=6cm,width=12cm,file=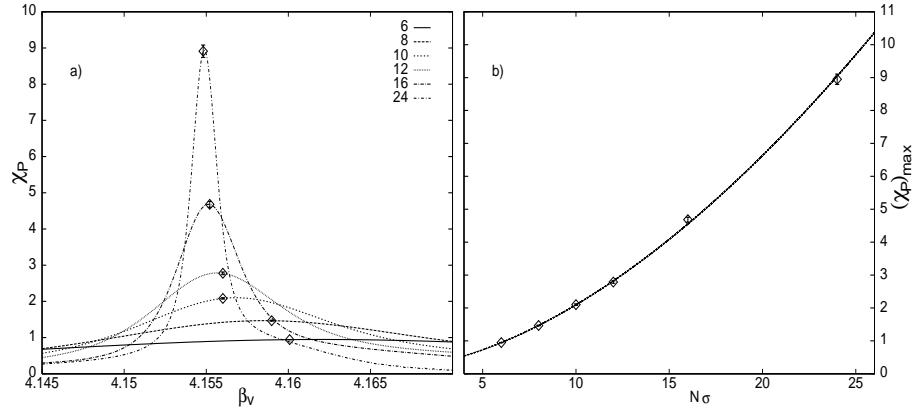}
\caption{a)The plaquette susceptibility, Eq.(\ref{eq.pksc}),
 as a function of \bv\ for \ns\ = 6, 8, 10, 12, 16 and 24. 
The points show the couplings where runs were made for each lattice;
the lines are Ferrenberg-Swendsen extrapolation.
b) Peak of the plaquette susceptibility plotted against \ns\ . 
Also shown is the best power law fit, discussed in the text. }
\label{fg.pkt}
\end{center}\end{figure}

Further insight into the nature of the transition can be gained by 
looking at the plaquette- and $L_a$-susceptibilities
\begin{eqnarray}
\label{eq.pksc}
\chi_P &=& 12\ns^3(\langle P^2 \rangle - \langle P \rangle ^2), \\
\label{eq.plsc}
\chi_{L_a} &=& \ns^3 (\langle L_a^2 \rangle - \la ^2).
\end{eqnarray}
For a first order bulk transition, the plaquette susceptibility is
expected to scale as the volume of the lattice \cite{cha}. Therefore,
for our \nt\ = 2 lattices, they are supposed to scale as $\ns^3$. The
peak heights are plotted against \ns\ in Fig. \ref{fg.pkt} b). Also
shown is the best fit to a form $\chi_{P_{max}}(\ns)=a+b~\ns^\omega$.
The corresponding values of the parameters are tabulated in table 
\ref{tb.fit}. The exponent is clearly different from the expectations 
for first order transition. 

\begin{table}
\caption{Fit parameters}
\vskip2mm
\begin{center}
\begin{tabular}{|c||c|c|c|c|}
\hline
 &$a$ &$b$ &$\omega$ &${\chi^2 \over d.f.}$ \\
\hline
\hline
$\chi_P$ &$.14 \pm .07$ &$(3.6^{+.8}_{-.7}) \times 10^-2$ &$1.73 \pm .07$ 
&1.12 \\
\hline
$\chi_{L_a}$ &$-3.6 \pm .6$ &$2.9^{+.6}_{-.5}$ &$1.78 \pm .07$ &.69 \\
\hline
\end{tabular}
\end{center}
\label{tb.fit}\end{table}	

\begin{figure}[htbp]
\begin{center}
\epsfig{height=6cm,width=12cm,file=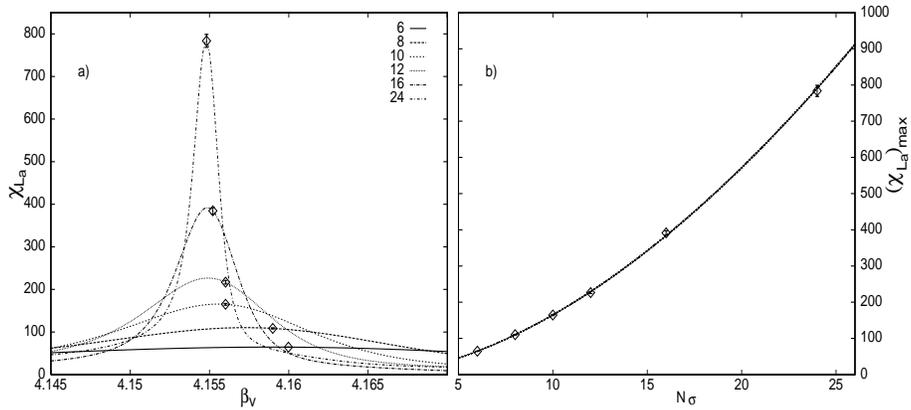}
\caption{a)The Polyakov loop susceptibility, Eq.(\ref{eq.plsc}), as 
a function of \bv\ for \ns\ = 6, 8, 10, 12, 16 and 24. 
b) Peak of the Polyakov loop susceptibility plotted against 
\ns\ . Also shown is the best fit discussed in text. }
\label{fg.pol}
\end{center}\end{figure}

In case of a deconfinement transition, the peak of the $L_a$
susceptibility may be expected to scale as $\ns^3$ for a first order
transition, and as $\ns^\omega$ with $\omega \sim 1.97$ for an Ising-like
transition. In Fig. \ref{fg.pol} b) the peak heights are
shown as a function of \ns\ . The best fit to the form
$a+b~\ns^\omega$ is also shown. It yields $\omega = 1.78 \pm 0.07$ 
which, amusingly, is not too far from expectations for an Ising-like 
transition. 

\begin{figure}[htbp]
\begin{center}
\epsfig{height=6cm,width=12cm,file=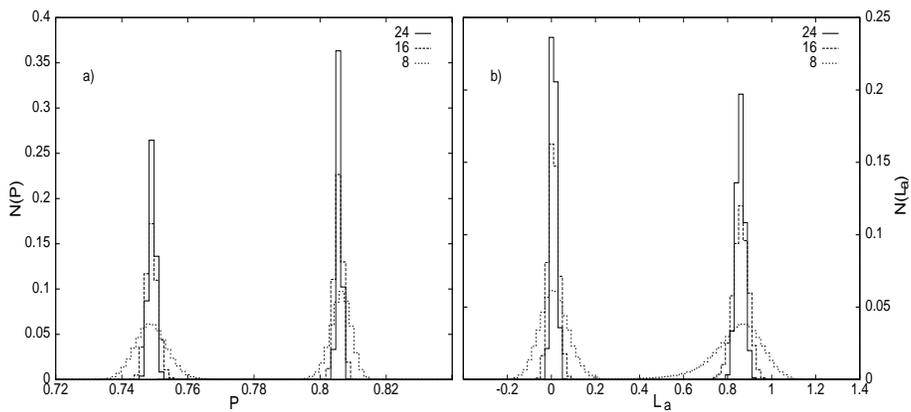}
\caption{Distribution of a) plaquette and b) $L_a$ for 
$\ns^3 \times 4$ lattices with \ns\ = 8, 16, 24.} 
\label{fg.nt4}
\end{center}\end{figure}

The above behavior is in contrast to what is seen for \nt\ = 4, 6 and 8 
lattices, where a strong first order transition was seen in ref.
\cite{sau2}. In fact, Fig. \ref{fg.hist} is to be contrasted 
with Figs. 6 and 7 of ref. \cite{sau2} for \nt\ = 4, where the peak 
positions remained stationary and the peak structure became sharper
with increase in \ns\ over the range 4 to 8.
We have checked that the behavior does not alter on increasing 
the spatial lattice size by making fresh simulations for \ns\ = 16 and
24 for \nt\ = 4.  The resulting histograms are shown in Fig.
\ref{fg.nt4}, along with the $8^3 \times 4$ data from ref. \cite{sau2}. 
They clearly indicate a first order transition for \nt\ = 4 lattices 
in the infinite volume limit, with discontinuities in plaquette and 
adjoint Polyakov loop given by, $\triangle P = .058 \pm 3, 
\triangle L_a = 0.87 \pm 0.04$. 

The second order transition for the \nt\ = 2 case, with an exponent
close to the 3-D Ising model, may prompt one to identify the transition
with the $SU(2)$ deconfinement transition, thereby vindicating the
shielding scenario above. However, on a closer examination, one sees
that the detailed properties of the transition are very different from
the $SU(2)$ case. The plaquette variable is smooth at the $SU(2)$ 
deconfinement transition point, and $\chi_{L_a}$ does not show
criticality there \cite{fin}, whereas from table \ref{tb.fit} one finds that 
both $\chi_{L_a}$ and $\chi_{P}$ diverge at the $SO(3)$ transition with 
essentially the same exponent. This indicates that here too one is seeing 
an intricate interplay of bulk and deconfinement transition effects, very
much like the higher \nt\ cases. We believe that the reason the order of the 
transition changes in going from higher \nt\ to \nt\ = 2 may simply be
the fact that the latter have effectively a smaller bulk-dimensionality. 
This interpretation, suggested by the near equality of the exponents for 
$\chi_{L_a}$ and $\chi_P$, indicates a more complicated interplay of the 
bulk and deconfinement transitions than the simple shielding scenario since 
a shielding is easy to justify only in the vicinity of a strong first order 
transition, while the \nt\ = 2 lattices show a mixing of effects of
bulk and deconfinement transitions even in the presence of a continuous
bulk transition.  Moreover, it also calls for a different approach to
disentangle the bulk effects from the deconfining ones.  In the next
section, we present results from an attempt in that direction.

\section{PHASE TRANSITION FOR \\ EXTENDED ACTION}
\label{sc.mono}

Since the $SO(3)$ bulk transition is known to be caused by a condensation of 
$Z_2$ monopoles, which are purely topological lattice artifacts and do not 
survive in the continuum limit, one may hope to get a clearer view of the 
physics of the deconfinement transition in the $SO(3)$ theory by suppressing 
them, without missing any important continuum physics.  One may be able
to address important physics questions associated with the
deconfinement transition in this manner, specially since there is no
obvious order parameter in this case.  The suppression 
can be achieved smoothly by adding a term to the action which
disfavors the monopole configurations. The most obvious choice is
to add a monopole chemical potential term to the action
\cite{hal2}, as considered below.

\subsection{Action and its properties}
\label{ssc.monoac}

We study the action \cite{hal2}
\beq
S_M (U, \sigma) = {\bv \over 2} \sum_p {\rm Tr}_f (U_p). ~\sg + \lm
\sum_c \sm~~,~~
\label{eq.mnac}
\eeq
where \sm\ is defined in Eq. (\ref{eq.mono}). 

\begin{figure}[htbp]
\begin{center}
\epsfig{height=6cm,width=7cm,angle=0,file=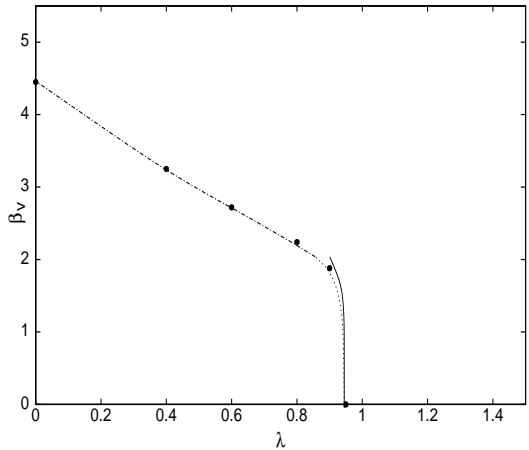}
\caption{The phase diagram for action \protect(\ref{eq.mnac}). The
dotted line denotes a second order transition while the dot-dashed
line denotes a first order line. The points are the actual simulation
points. The full line is the approximation of Eq. \protect(\ref{eq.eff}).}
\label{fg.phase}
\end{center}\end{figure}

This action has a line of phase transitions in the (\bv\ , \lm\ ) 
plane \cite{hal2}. For \lm\ = 0, it reduces to the action of 
Eq. (\ref{eq.vlac}), with a first order transition at $\bvc
\sim 4.5$. For \bv\ = 0, the theory is dual to four dimensional Ising
model with coupling $\tilde{\lm} = {1 \over 2} {\rm ln} ~{\rm coth}
~\lm$ \cite{sav}. It therefore has a second order phase transition, 
at $\lm_c \sim .95$. For small \bv\ , one can integrate out the gauge
field : this contributes, in the leading order, to only a
renormalization of the coupling
\beq
\lm \to \lm_{eff} = \lm + 4 \bt_{\frac {1} {2}}^6
\label{eq.eff}
\eeq
where $\bt_{\frac{1}{2}}$ is given in eq. (\ref{eq.strong}).
Therefore, the transition will remain second order, at least to a
point where the leading order approximation is valid. As \bv\ is
increased, it becomes first order and joins the transition
point on the \bv\ - axis. We show the bulk phase diagram in
Fig. \ref{fg.phase}.  Apart from some details like the second order line 
etc., it is qualitatively the same as that in ref.\cite{hal2}.  For  large 
\bv\ , the large \lm\ region is smoothly connected to the small 
\lm\ region. For, due to Eq. (\ref{eq.weak}) the second term in action
(\ref{eq.mnac}) is irrelevant here (any other configuration is
strongly suppressed and can be treated in the dilute gas approximation). 
So the weak coupling limit is not changed by the addition of this
term.

\subsection{Study of the Finite Temperature Phase Transition}
\label{ssc.monors}

In order to study the true deconfinement phase transition
which is unaffected by the bulk effects, we need to 
choose a value of $\lm > \lm_c$, so that the bulk transition can be 
avoided. For our simulations, we have used \lm\ = 1. We
checked that with this value of \lm\, the plaquette variable is
indeed smooth everywhere, indicating the lack of any bulk phase
transition. The approximate location of the deconfinement transition
is indicated by a steep rise in \la\ . We studied the deconfinement transition
for lattices with \nt\ = 2 and 4. For finite size analysis, lattices
with \ns\ = 8, 12 and 16 were used for both \nt\ .

For \nt\ = 2, the deconfinement transition
was found to be at $\bv \sim 1.9$, while it was found to be at
\bv\ around 2.3 for the \nt\ = 4 lattices. Both of these are close to the
corresponding transition points for pure $SU(2)$ and far from the
transition points for action (\ref{eq.vlac}) (see Sec. \ref{ssc.nt2rs}).
We carried out simulations at a few \bv\ around the approximate transition 
point for each lattice size and then used multi-histogramming 
methods \cite{fer2} to obtain the physical 
variables in the whole critical region. The details of the
simulation are given in Appendix \ref{app.sim}. In
Fig. \ref{fg.monola} the behavior of \la\ across the transition
point, as obtained by the multi-histogramming extrapolation, is shown. 
\la\ is seen to rise smoothly without any signs of a
discontinuity in it.  This is similar to its known behavior across
the $SU(2)$ deconfinement transition. We also found that the 
the \la\ susceptibility did not show any peak at the transition 
point, again in agreement with the expectations from the $SU(2)$ theory. 

\begin{figure}[htbp]\begin{center}
\epsfig{height=6cm,width=6cm,figure=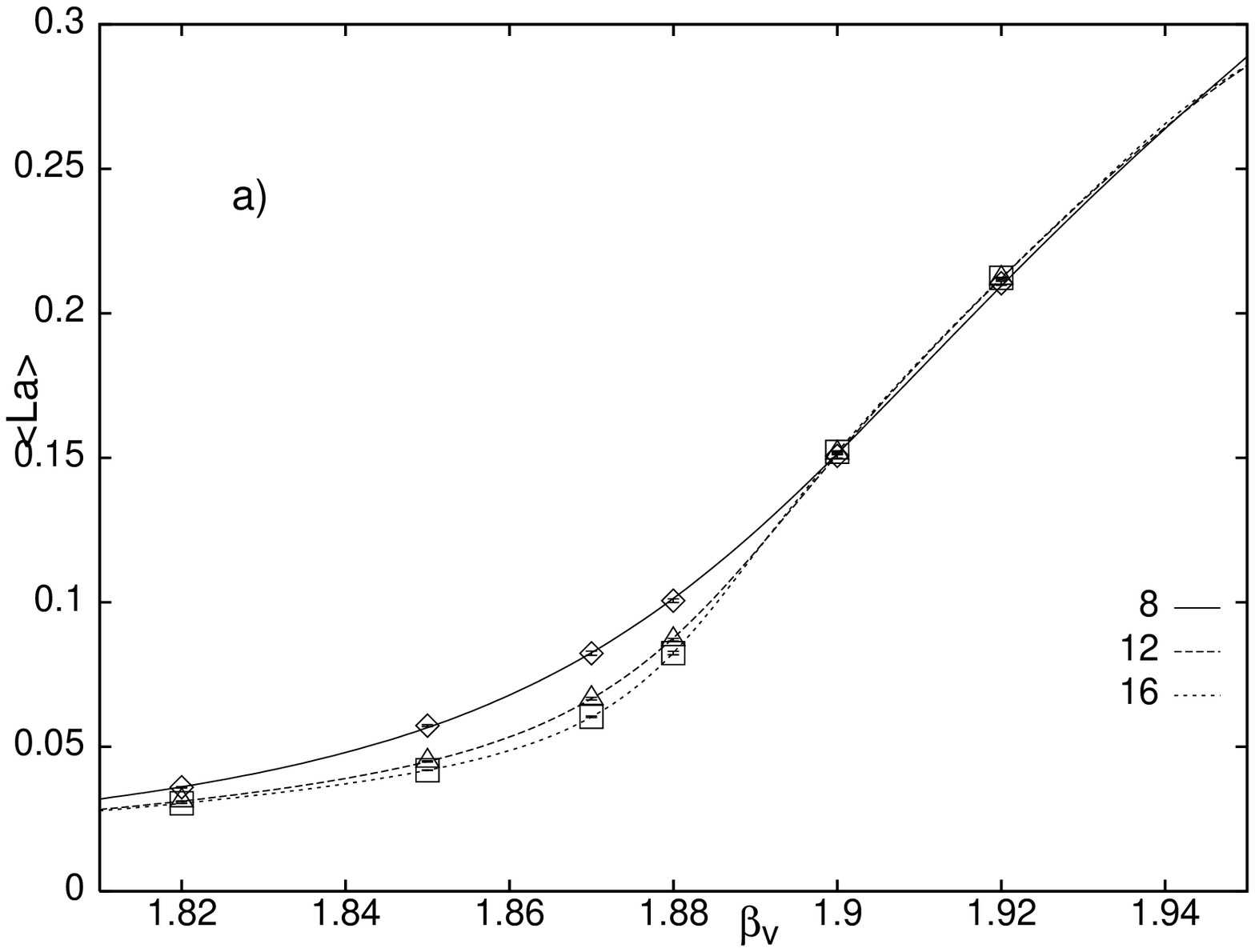}
\epsfig{height=6cm,width=6cm,figure=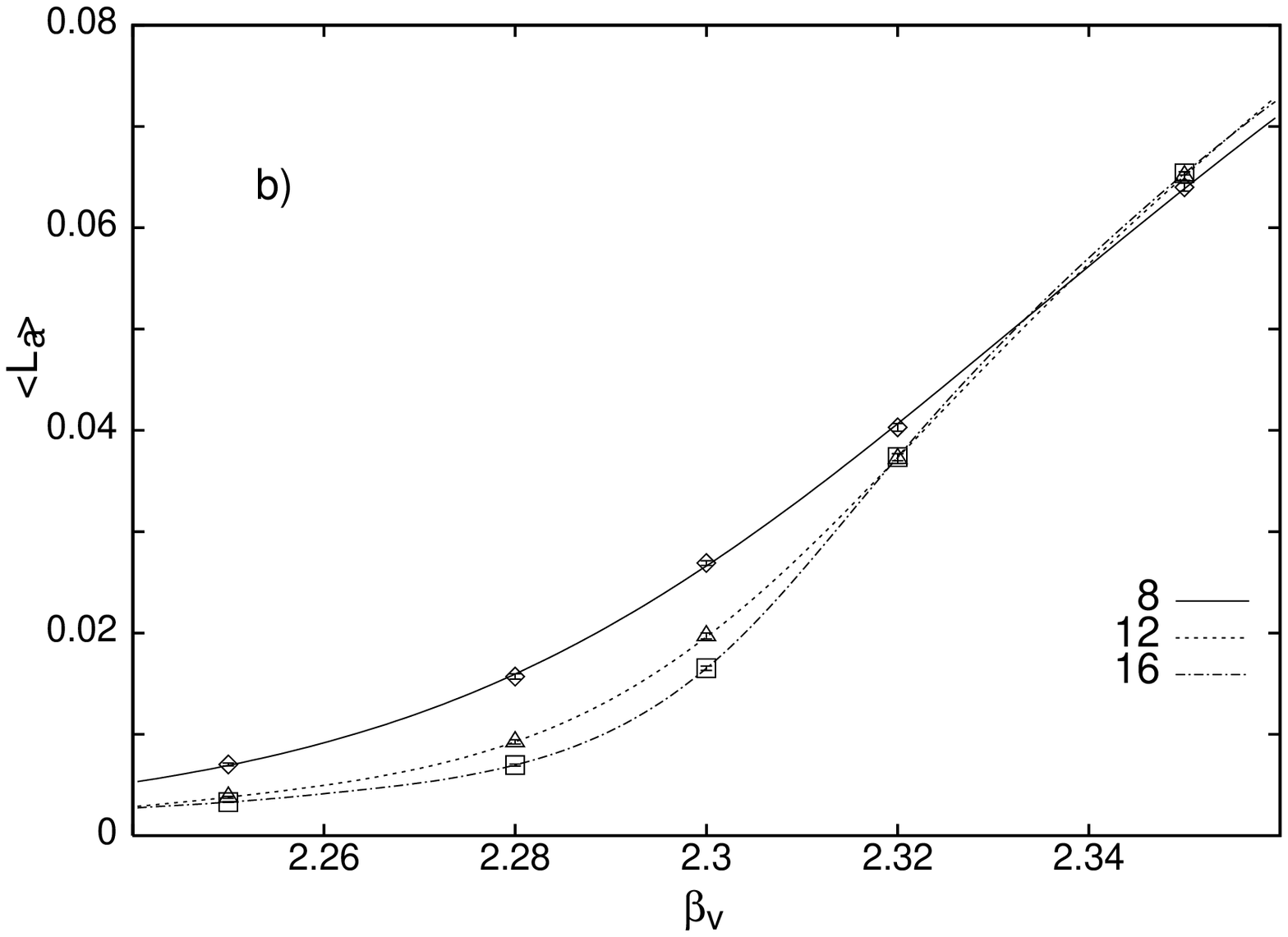}
\caption{a) \la\ across the deconfining point for \nt\ = 2 lattices
with \ns\ = 8, 12, 16. The points with error bars show the actual
couplings where simulations were made. The curve is obtained by using
multi-histogramming methods. b) Same for \nt\ = 4.}
\label{fg.monola}\end{center}\end{figure}

\begin{figure}[htbp]
\begin{center}
\epsfig{height=6cm,width=6cm,file=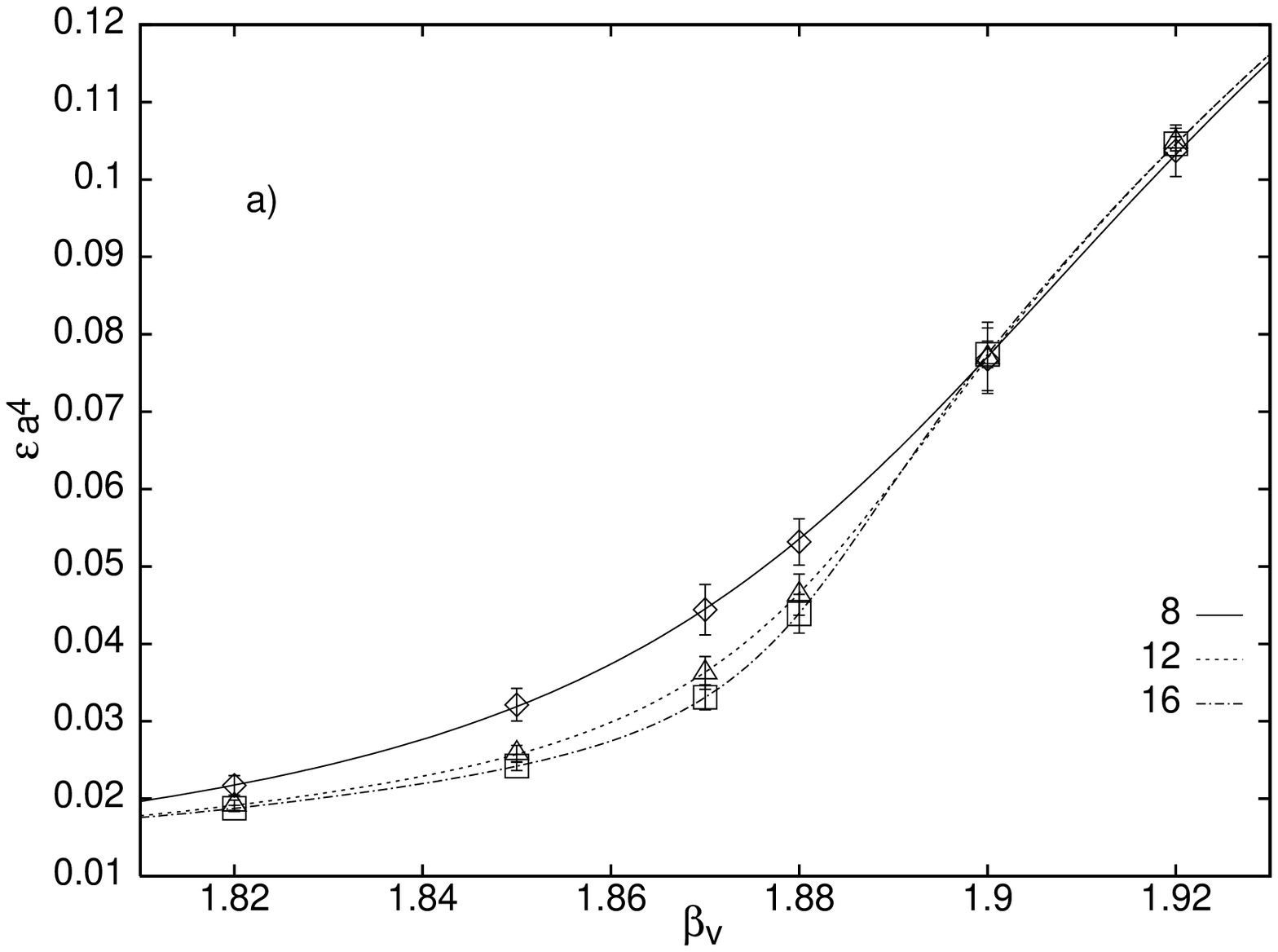}
\epsfig{height=6cm,width=6cm,file=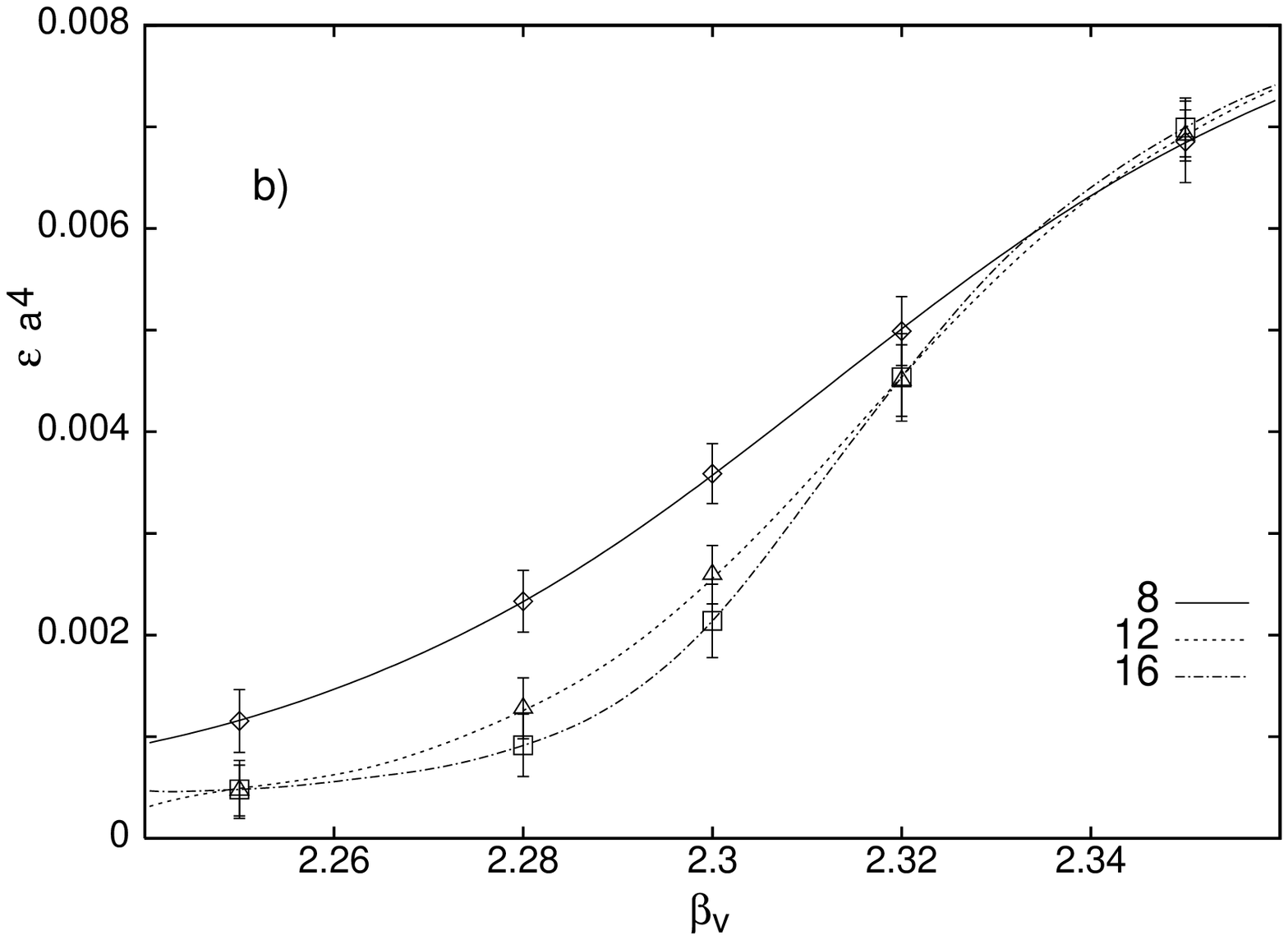}
\caption{The energy density, Eq. (\ref{eq.eneg}), in the critical region
of a) $\ns^3 \times 2$ and b) $\ns^3 \times 4$ lattices for \ns\ = 8, 12,
16.} 
\label{fg.energ}
\end{center}\end{figure}

A quantitative study of the order of the deconfinement transition is
made difficult by the absence of an order parameter for this transition. 
The two physical variables that can be used are the energy density and
specific heat. Extraction of these observables from lattice simulations 
is, however, difficult, as they are not simply related to objects directly
measured in simulations. Since our purpose
is only to show the existence of a deconfinement transition, and not
to make quantitative predictions about the transition, we use for
these variables the approximate expressions (see Appendix \ref{app.form})
\begin{eqnarray}
\label{eq.eneg}
\varepsilon a^4 &=& 3 \bv \langle P_\tau - P_\sigma \rangle, \\
\label{eq.spht}
C_v a^3 &=& 3 \nt N_p \bv^2 {\rm var} (P_\tau - P_\sigma),
\end{eqnarray}
where $N_p$ is the number of spatial plaquettes,
$P_\tau$ and $P_\sigma$ are spatial and temporal plaquette variables
respectively, and ${\rm var}(x)=\langle x^2 \rangle - \langle x
\rangle ^2$. The energy density for \nt\ = 2 and 4 lattices is shown 
in Fig. \ref{fg.energ}. It is seen to rise significantly across the 
transition for both the \nt\ values, justifying the
identification of the transition as a deconfining one. 
The rise becomes steeper with increase in the spatial volume,
indicating a true transition as opposed to a crossover. 

\begin{figure}[htbp]
\begin{center}
\epsfig{height=6cm,width=8cm,file=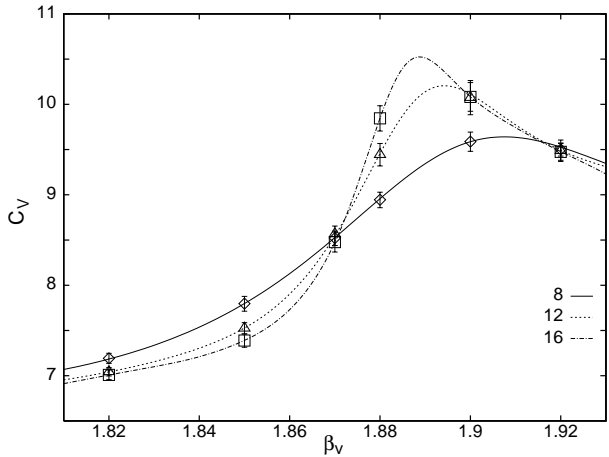}
\caption{The specific heat density, Eq. (\ref{eq.spht}), in the critical 
region of $\ns^3 \times 2$ lattices for \ns\ = 8, 12, 16.}
\label{fg.spht}
\end{center}\end{figure}

The behavior of the specific heat for the \nt\ = 2 lattices 
is shown in Fig. \ref{fg.spht}.
The position of the peak corresponds to the critical point. The
peak height is seen to increase with the lattice size, and the peaks
become sharper. Both these behaviors are expected for a 
deconfinement phase transition. While our expressions for specific heat,
Eq. (\ref{eq.spht}), are somewhat approximate to obtain an accurate critical 
exponent, it is interesting to note that a linear fit to the peak
heights gives ${\alpha \over \nu} = 0.13 \pm 0.04$, which is consistent 
with the three dimensional Ising model value ${\alpha \over \nu} 
\simeq .17$. For the \nt\ = 4 lattices, the energy density being 16
times smaller in lattice units, the resultant specific heat curves
were very noisy and inconclusive; a measurement of the exponent here
is computationally very demanding and beyond our reach.

\begin{figure}[htbp]
\begin{center}
\epsfig{height=6cm,width=7cm,file=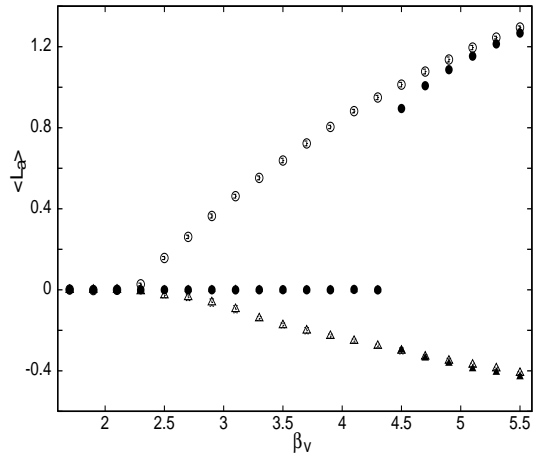}
\caption{\la\ for a range of couplings, showing the two different
states in the high temperature phase. The filled circles and triangles
correspond to the Villain action of Sec. \protect\ref{sc.nt2} while
the open circles and triangles correspond to the monopole suppressed
action.}
\label{fg.negla}
\end{center}\end{figure}

Thus the deconfinement transition for the monopole suppressed action of
$SO(3)$ is very similar in nature to that for the $SU(2)$ Wilson action,
although the former has no order parameter.  A significant difference in 
the two, however, is the presence of negative \la\ states in the
$SO(3)$ theory.   For the $SU(2)$ Wilson action, \la\ approaches a unique 
positive value $\to 3 ~{\rm as}~ \f \to \infty$ in the high temperature 
phase. However, for $SO(3)$, \la\ manifests itself in two states in the 
high temperature phase. One is like the high temperature state of $SU(2)$ 
but the other has a negative \la\ , which approaches $- 1$ as $\bv \to
\infty$. In ref. \cite{sri}, it was proposed that the negative \la\ state
corresponds to a new bulk phase which is separated from the $\la \sim 0$
bulk phase by a bulk phase transition.  From the behavior of the correlation
functions, and the behavior of \la\  in presence of a ``magnetic'' term
$h L_a$ in the action, it was concluded in ref. \cite{sau2} that 
both the positive and negative \la\ states correspond to the same physics. 
In the monopole-suppressed action, there is no bulk transition.
However, we still find the negative \la\ state, very similarly to the
monopole unsuppressed action. In Fig. \ref{fg.negla} we show the
behavior of \la\ for a large range of the coupling \bv\ , for both 
\lm\ = 0 (the pure Villain action) and \lm\ = 1 (the monopole suppressed 
case). Other than the shift in transition point, there is no essential
difference between the two.  Therefore the possibility advocated in 
ref. \cite{sri} of the negative \la\ state being another bulk phase, 
separated by a bulk transition, is disfavored. 

\section{Summary and Discussion}
\label{sum}

We studied the $SO(3)$ lattice gauge theory in a Villain form numerically.
Our main objective was to disentangle the effects of the bulk
phase transition in this theory from those of the deconfinement phase
transition in it, if any, and then to compare the latter with the $SU(2)$
theory which is expected to be in the same universality class.  We attempted 
this by investigating the case of smaller \nt\ = 2 lattices and later by 
suppressing the $Z_2$ monopoles.  Going to \nt\ = 2, the nature of the only 
transition observed was seen to change substantially compared to the 
earlier \nt\ = 4-8 studies. The transition point had a visibly large 
shift, and the order of the transition was found to change to second. 
However, the curious mix of bulk and
finite temperature transitions, seen on lattices with higher \nt\ ,
continued, as the behavior of the plaquette variables and that of
\la\ , the adjoint Polyakov loop, were very similar. Determination of
the critical exponents for their respective susceptibilities suggested a 
coincident bulk and deconfinement phase transition, similar to the
higher \nt\ studies but with a second order bulk transition. The 
scenario of bulk transition shielding the deconfinement transition, 
which was a plausible one for the higher \nt\ lattices, is therefore 
disfavored, indicating the necessity of a more complicated scenario.

In order to get a clearer view of deconfinement in the $SO(3)$ theory,
the bulk transition was evaded by adding a term in the action to
suppress the topological objects  driving the bulk transition. The
resulting theory was found to have a deconfinement transition whose nature 
is very similar to the deconfinement transition in $SU(2)$. This is very
interesting as the center symmetry, which plays a crucial role in the 
$SU(2)$ deconfinement transition, is trivial here, and the theory does
not have any order parameter for its deconfinement transition. 
The location of the transition is also seen to be very different from that
where the coincident bulk/deconfinement transition was seen for the Villain
action without monopole suppression. It seems, therefore,
that the bulk transition affects the deconfinement transition in a
very subtle and nontrivial way, making it always coincide with itself. It
would be very interesting to see two separate 
transitions for the original $SO(3)$ theory without any monopole
suppression on a big enough lattice, since these monopoles are mere
lattice artifacts.

\section{Acknowledgments}
One of us (S.D.) would like to thank Profs. Frithjof Karsch, Sumit Das
and Sourendu Gupta, and the other (R.V.G.) would like to thank 
Dr. Manu Mathur for stimulating discussions and suggestions.

\appendix
\section{Details of Numerical Simulations}
\label{app.sim}
In this appendix we give details of the numerical simulations 
mentioned in the text. For all the simulations, we have used heat bath
algorithms for both the gauge variables and the \sg\ . An iteration
consisted of updating all the link variables once, followed by an
update of all \sg\ . For the \nt\ = 2 measurements of 
Sec. \ref{ssc.nt2rs}, the transition point was approximately located for
each lattice by looking for a phase coexistence. At this point a small
run was made, and the spectral density method \cite{fer1} was used to
locate the susceptibility peak where a long simulation run
was made. We used 1.6 million configurations for each lattice. Table
\ref{tb.mc} lists the couplings at which the long runs were made for
different \ns\, and the autocorrelation times for plaquette and $L_a$. 
For the extrapolations to nearby couplings using the spectral density
method, the Jackknife estimate was used for both the extrapolated
value and the error, using 20 blocks for the analysis.
For the histograms of $16^3 \times 4$ and $24^3 \times 4$ lattices of
this section, we produced 50000 configurations for each lattice,
which was sufficient to clearly establish the peaks and nature of the 
histograms. The phase coexistence region was much broader for these 
lattices; we took $\bv=4.45$, approximately at the middle of the
region, for the longer runs. Due to the strong first order nature of the 
transition, tunnelings are not seen and we had to combine configurations from
two runs to get the peaks.

\begin{table}
\caption{Details of the Monte Carlo simulations of Sec. 
\protect\ref{ssc.nt2rs}.}
\vskip2mm
\begin{center}
\begin{tabular}{|c|c|c|c|}
\hline
\ns\ &\bv\ &$\tau_P$ &$\tau_{L_a}$ \\
\hline
\hline
6 &4.16 &516.8 &754.7 \\
\hline
8 &4.159 &1010.4 &1294.3 \\
\hline
10 &4.156 &1738.5 &1878.8 \\
\hline
12 &4.156 &2245.7 &2254.6 \\
\hline
16 &4.1552 &2477.6 &3896.5 \\
\hline
24 &4.1548 &3155.8 &4753.1 \\
\hline
\end{tabular}
\end{center}
\label{tb.mc}\end{table}

For the simulations in Sec. \ref{ssc.monors}, it was noticed that 
the plaquette variables have an extraordinarily large autocorrelation
time. This can be reduced dramatically by implementing the
energy-conserving steps corresponding to the center symmetry
transformation (Eq. (\ref{eq.center})). After every combined step
of the previous section, we added another sweep where a fixed fraction
(arbitrarily chosen to be one-fourth) of the total number of links, 
randomly chosen, are flipped along with all the plaquettes touching 
the link. An iteration consisted of the combination of this sweep 
with the combined heat bath step, and measurement was taken after every
iteration. It was checked that for all the runs, the autocorrelation
time was less than 100. For the \nt\ = 2 results of this section, from
the rise in \la\ the critical point was estimated to be $\stackrel{<}
{\sim} 1.90$. We ran simulations at \bv\ = 1.82, 1.85, 1.87, 1.88,
1.90 and 1.92, producing half million configurations at each coupling
for each lattice. For the \nt\ = 4 case, the critical point was at
$\sim 2.30$. We produced one million configuration at \bv\ = 2.30, and half
million configurations each at \bv\ = 2.25, 2.28, 2.32 and 2.35. 
For interpolating to other values of \bv\ , each dataset was divided
into very finely meshed histograms, and the multihistogramming techniques
of ref. \cite{fer2} was applied over the whole set. The central values
and the errors quoted are Jackknife estimates, 20 blocks being used
for the analysis in each case.

\section{The Energy Density and Specific Heat}
\label{app.form}

The expressions for energy density and specific heat density, Eqs. 
(\ref{eq.eneg}) and (\ref{eq.spht}), can be easily obtained using the
methods of ref.\cite{eng}. We outline here the derivation of these results,
emphasizing the approximations involved.

For finite temperature theory, it is necessary to take into account 
the asymmetry in the spatial and
temporal directions by using two different couplings. Writing
\beq
S = \bt_\sigma \sum_{{\rm spatial} ~p} P_\sigma + \bt_\tau \sum_{{\rm
temporal} ~p} P_\tau,
\label{eq.asymac}
\eeq
the correct continuum limit is recovered by writing $\bt_\sigma = 
{4 \over g_\sigma^2} 
{a_\tau \over a_\sigma}$ and $\bt_\tau = {4 \over g_\tau^2} 
{a_\sigma \over a_\tau}$, where $a_\tau$
and $a_\sigma$ are the temporal and spatial lattice spacings,
respectively. The couplings $g_\sigma$ and $g_\tau$ both equal
the usual bare coupling $g$ when the two lattice spacings are equal. 
In general, one can expand them perturbatively in g near the continuum
limit \cite{eng}: \[ g^{-2}_{\sigma, \tau} (a_\sigma, a_\tau) 
= g^{-2} (a_\sigma) + c_{\sigma, \tau} (a_\sigma / a_\tau) + O(g^2). \]

Defining the inverse temperature $\bt = \nt a$, the energy density 
is given by
\begin{eqnarray}
\varepsilon &=& - {1 \over V} {\partial \over \partial \bt} {\rm ln} Z
|_V \nonumber \\
&=& - {1 \over \ns^3 a_\sigma^3 \nt} \Bigl\langle {\partial S \over
\partial a_\tau} |_{a_\sigma} \Bigr\rangle.
\label{eq.engder}
\end{eqnarray}
 From Eq. (\ref{eq.asymac}), one sees that evaluation of $\partial S /
\partial a_\tau |_{a_\sigma}$ requires knowledge of the
nonperturbative beta functions $\partial g^{-2}_{\sigma, \tau} /
\partial a_\tau |_{a_\sigma}$, which, however, are not known for $SO(3)$
gauge theory. One can, therefore, use the perturbative expressions for
the same, using the expansion in terms of $g^{-2}$ given above. The
perturbative beta functions are exactly similar to the $SU(2)$ case of 
ref. \cite{eng}. The contribution from this term is found to be small
and does not affect the critical behavior. Ignoring this term, and 
putting $a_\sigma = a_\tau = a$ after taking the derivatives in
Eq. (\ref{eq.engder}, one gets the expression (\ref{eq.eneg}) 
for the energy density.  

Similarly the specific heat at fixed volume is
\begin{eqnarray}
C_v &=& {\partial \varepsilon \over \partial T}|_V = - \bt^2 
{\partial \varepsilon \over \partial \bt} \nonumber \\
&=& {\bt^2 \over \nt^2 V} \Bigl\{ \Bigl\langle {\partial^2 S \over
\partial a_\tau^2 } |_{a_\sigma} + \bigl({\partial S \over \partial
a_\tau} |_{a_\sigma} \bigr)^2 \Bigr\rangle - \Bigl\langle {\partial S \over
\partial a_\tau} |_{a_\sigma} \Bigr\rangle ^2 \Bigr\} \nonumber \\
&=& {1 \over V} \Bigl \{ 2 \beta N_p \langle P_\tau - P_o \rangle + \bt^2
N_p^2 {\rm var} \bigl( P_\tau - P_\sigma \bigr) \Bigr \},
\label{eq.sphtder}
\end{eqnarray}
where $P_o$ is the plaquette variable on a symmetric lattice,
and the coupling derivative terms have again been neglected. 
The first term in Eq. (\ref{eq.sphtder}) is of same order as 
the neglected derivative terms,
and should be omitted for consistency, giving us the expression
(\ref{eq.spht}). We have also checked that inclusion of the first term
does not change the results in any significant way.

\end{document}